\documentclass[twocolumn,showpacs,preprintnumbers]{revtex4}
\usepackage{amsmath}
\usepackage{amssymb}
\usepackage{graphicx}
\usepackage{bm}
\begin{document}

\title{Symmetry Energy Effects on Fusion Cross Sections}

\author{C.Rizzo $^{a,b}$,V.Baran$^{c}$, M. Colonna$^{a}$, 
 A. Corsi$^{d}$, M.Di Toro$^{a,b,*}$}

\affiliation{
        $^a$LNS-INFN, I-95123, Catania, Italy\\
	$^b$Physics and Astronomy Dept. University of Catania, Italy\\
        $^c$NIPNE-HH, Bucharest and Bucharest University, Romania\\
        $^d$Physics. Dept., University of Milano and INFN Sez. Milano\\
        $^*$ email: ditoro@lns.infn.it}

\begin{abstract}

We investigate the reaction path followed by Heavy Ion Collisions with exotic 
nuclear beams at low energies. We will focus on the interplay between reaction
mechanisms, fusion vs. break-up (fast-fission, deep-inelastic), that in exotic 
systems is expected to be influenced by the symmetry energy term at densities 
around the normal value. 
The evolution of the system is described by a Stochastic Mean Field transport 
equation (SMF),  where two parametrizations for the density 
dependence of symmetry energy (Asysoft and Asystiff) are implemented, 
allowing one to explore the sensitivity
of the results to this ingredient of the nuclear interaction.  
The method described here, based on the event by event evolution of phase 
space quadrupole collective modes will nicely allow to extract the fusion 
probability at relatively early times, when the transport results are reliable.
Fusion probabilities for reactions induced by $^{132}$Sn on $^{64,58}$Ni 
targets at 10 AMeV are evaluated. We obtain larger fusion cross sections for 
the more n-rich
composite system, and, for a given reaction, 
in the Asysoft choice. 
Finally a collective charge equilibration mechanism (the Dynamical Dipole) 
is revealed in both fusion and break-up events, depending on the 
stiffness of the symmetry term just below saturation.

\end{abstract}

\maketitle
\date{\today}

\section{Introduction}

Production of exotic nuclei has opened the way to explore, in 
laboratory conditions, 
new aspects of nuclear structure and dynamics
up to extreme ratios of neutron (N) to proton numbers (Z). 
An important issue addressed
is the density dependence of the symmetry energy term in the nuclear
Equation of State (EOS), of interest also for the properties of
astrophysical objects  \cite{bao01,ste05,bar05a,bao08}. 
By employing Heavy Ion Collisions (HIC), at
appropriate beam energy and centrality, the
isospin dynamics at different densities 
of nuclear matter can be investigated
\cite{bar05a,bao08,tsa_epj,tsa09,bar04,Kel10,fil05}.
   
In this work we will focus the attention on the interplay of fusion vs. 
deep-inelastic mechanisms  for dissipative HIC 
with exotic nuclear beams at low energies, just above the Coulomb Barrier
(between $5$ and $20$ AMeV), where unstable ion beams with large asymmetry
will be soon available.  
We will show that the competition between reaction mechanisms can be used 
to study properties of  the symmetry energy term in a  density range
around the normal value. 
Dissipative collisions at low energy are characterized by  
interaction times that are quite long and by a large coupling among 
various mean field modes that may eventually lead to the break-up of the 
system. 
Hence the idea is to probe how the symmetry energy will influence such 
couplings 
in neutron-rich systems with direct consequences on the fusion probability.
We will show that, 
within our approach, the reaction path is fully characterized by the 
fluctuations, at suitable time instants, of phase space quadrupole collective
modes that lead the composite system either to fusion or to break-up.

Moreover, it is now well established that in the same energy range, for 
dissipative reactions
between nuclei with different $N/Z$ ratios, the charge equilibration process 
has a collective character resembling
a large amplitude Giant Dipole Resonance (GDR), see the recent \cite{bardip09}
and refs. therein.
The gamma yield resulting from the decay of such pre-equilibrium
isovector mode can encode information about the early stage of the 
reaction \cite{cho93,bar96,sim01,bar01b,bar01}. 
This collective response is appearing in the 
intermediate neck region, while the system is still in a 
highly deformed dinuclear configuration with large surface contributions, 
and so it will
be sensitive to the density 
dependence of symmetry
energy below saturation \cite{bardip09}. Here we will show that this mode
is present also in break-up events, provided that a large dissipation is 
involved. 
In fact we see that the strength of such fast dipole emission is not much 
reduced passing from fusion to very deep-inelastic mechanisms. This can 
be expected  
from the fact that such excitation is related to an entrance channel 
collective oscillation.
Thus we suggest the 
interest of a study of the prompt gamma radiation, with its characteristic 
angular anisotropy \cite{bardip09}, even in deep-inelastic
collisions with radioactive beams.

The paper is organized as follows. In Sect.II we present our transport 
approach to the low energy HIC dynamics with description of the used 
symmetry effective potentials. Sect.III is devoted to the analysis 
of $^{132}Sn$ induced reactions
with details about the procedure to select fusion vs. break-up events. 
In Sect.IV we
discuss symmetry energy effects on the competition between fusion and 
break-up (Fast-fission, Deep-inelastic, Ternary/Quaternary-fragmentation)
mechanisms. The dependence
on symmetry energy of the yield and angular distribution of the 
Prompt Dipole Radiation, expected for entrance channels with large charge 
asymmetries, is presented in Sect.V. Finally in Sect.VI we summarize 
the main results and we suggest some experiments to be performed at the 
new high intensity 
Radioactive Ion Beam (RIB) facilities in this low energy range.

\section{Reaction Dynamics}



The reaction dynamics is described
by a Stochastic Mean-Field (SMF) approach, extension of the microscopic
Boltzmann-Nordheim-Vlasov transport equation \cite{bar05a}, where
the time evolution of the semi-classical 
one-body distribution function $f({\bf r},{\bf p},t)$ is 
following
a Boltzmann-Langevin evolution dynamics (see \cite{rizzo_npa} 
and refs. therein): 
\begin{equation}
\frac{\partial f}{\partial t}+\frac{\bf p}{m}\frac{\partial f}
{\partial {\bf r}}+
\frac{\partial U}{\partial {\bf r}}\frac{\partial f}{\partial 
{\bf p}}=I_{coll}[f]+\delta I[f].
\end{equation}
In the SMF model the fluctuating term $\delta I[f]$ 
is implemented in an approximate way, through stochastic 
spatial density fluctuations \cite{Mac_npa}.
 Stochasticity is essential to get distributions, as well as to allow the 
growth of dynamical instabilities.
In order to map the particle occupation at each time step,
gaussian phase space wave packets (test particles) are considered. 
In the simulations 100 test particles per nucleon have been employed for an 
accurate description of the mean field dynamics. 
In the collision integral, $I_{coll}$, an 
in-medium depending  
nucleon-nucleon cross section, via the local density, is employed
\cite{li93}. 
The cross section is set equal to zero for nucleon-nucleon 
collisions below 50 MeV of relative energy. In this way we avoid spurious 
effects, that may dominate in this 
energy range when the calculation time becomes too large.
In spite of that, for low energy collisions, the simulations cannot be 
trusted on the time scale
of a compound nucleus formation, mainly for the increasing numerical noise.
As it will be explained in Section III.B,
the nice feature of the procedure described here to evaluate the fusion 
probability is that, on the basis of a shape analysis in phase space,  
we can separate fusion and break-up trajectories at rather early times, 
of the order of 200-300 fm/c,
when the calculation can still be fully reliable. 

The mean field is built from Skyrme forces: 
\begin{eqnarray}
U_{n,p}&=&A\frac{\rho}{\rho_0}+B(\frac{\rho}{\rho_0})^{\alpha+1}
+C(\rho)
\frac{\rho_n-\rho_p}{\rho_0}\tau_q+ \nonumber  \\
&+&\frac{1}{2} \frac{\partial C}{\partial \rho} \frac{(\rho_n-\rho_p)^2}
{\rho_0}
\end{eqnarray}
where $q=n,p$ and $\tau_n=1, \tau_p=-1$.
The coefficients $A,B$ and the exponent $\alpha$, characterizing the 
isoscalar part of
the mean-field, are fixed requiring that the saturation properties of
 symmetric 
nuclear matter ($\rho_0=.145fm^{-3}$, $E/A=-16MeV$), with a compressibility 
modulus around {\bf $200~MeV$},
are reproduced.
The function $C(\rho)$ will give the potential part of the symmetry energy:
\begin{equation}
\frac{E_{sym}}{A}(\rho,T=0) = 
\frac{E_{sym}}{A}(kin)+\frac{E_{sym}}{A}(pot)\equiv 
\frac{\epsilon_F}{3} + \frac{C(\rho)}{2\rho_0}\rho
\end{equation}

\begin{figure}
\begin{center}
\includegraphics*[scale=0.33]{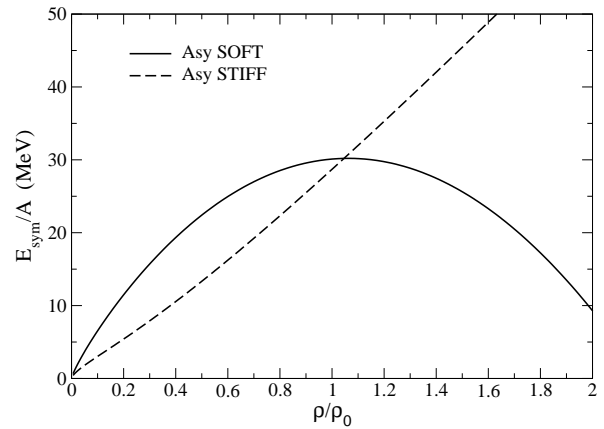}
\end{center}
\caption{Density dependence of the symmetry energy for the two
parametrizations. Solid line: Asysoft. Dashed line: Asystiff}
\label{esym}
\end{figure}
For the density dependence of the symmetry energy,
we have considered two different parametrizations \cite{col98,bar02},
that are presented in Fig.\ref{esym}.
In the Asysoft EOS choice, $\frac{C(\rho)}{\rho_0}=482-1638 \rho$, 
the symmetry energy has a weak density dependence close to the saturation, 
being 
almost flat around $\rho_0$. For the Asystiff
case, $\frac{C(\rho)}{\rho_0}=\frac{32}{\rho_0}\frac{2 \rho}{\rho+\rho_0}$, 
the symmetry energy is quickly decreasing for densities below normal 
density. Aim of this work is to show that fusion probabilities,
 fragment properties in break-up events, 
 as well as properties of prompt collective modes,   
in collisions induced by neutron-rich exotic beams, are sensitive to 
the different slopes of the symmetry term around saturation.



\section{Fusion dynamics for $^{132}Sn$ induced reactions}

In order to study isospin and symmetry energy effects on the competition 
between fusion and break-up (deep-inelastic) we consider the reactions
$^{132}Sn~+~^{64,58}Ni$ at 10 AMeV, having in mind that $^{132}Sn$ beams 
with good intensities in 
this energy range will be soon available in future Radioactive
Ion Beam facilities.
In particular, we have performed collision simulations for semi-peripheral 
impact parameters 
(from b =4.5 fm to b = 8.0 fm, with $\Delta$b= 0.5 fm), to explore the 
region of the 
transition from fusion to break-up dominance. 
The transport equations clearly give fusion events at central impact 
parameters and break-up events for peripheral collisions, but there are 
some problems when we consider semi-peripheral impact parameters at such 
low energies, since the time scales for break-up are not compatible with 
the transport treatment, as already noted.
It is then  not trivial to extract the fusion probability from the early 
dynamics of the system and test the sensitivity to the asy-EOS.  
Therefore we have tried to find a reliable criterion that can indicate 
when the reaction mechanism is changing, from fusion to deep-inelastic 
dominance.  This will also allow to evaluate the corresponding absolute 
cross sections.

The new method is based on a phase space analysis of quadrupole 
collective modes. 
The information on the final reaction path is deduced investigating the 
fluctuations of the system at early times (200-300 fm/c), 
when the formation of
composite elongated configurations is observed
and phenomena associated with surface metastabity and/or instability may 
take place. 
At later times, when the SMF dynamics is not reliable, 
the evolution of the most relevant degrees of freedom 
could be followed within a more macroscopic 
description, where the system is characterized in terms of global observables, 
for which the full treatment of fluctuations in phase space 
is numerically affordable \cite{Leonid}.   
However, we will show that a consistent picture of the fusion vs. 
break-up probabilities
can be obtained already from a simpler analysis of phase space 
fluctuations in the
time interval indicated above.

We start considering the time evolution, in each event,  
of the quadrupole moment in coordinate space which is given by: 
$$Q(t)=<2z^2(t)-x^2(t)-y^2(t)>,$$
averaged over the space distribution in the composite system. 
At the same time-steps we construct also the quadrupole moment in momentum 
space: 
$$QK(t)=<2p_z^2(t)-p_x^2(t)-p_y^2(t)>,$$
in a spatial region around the center of mass. 
The z-axis is along the rotating  
projectile-like/target-like direction, the x-axis is on the reaction plane.

\subsection{Average dynamics of shape observables}

We run 200 events for each set of macroscopic initial conditions and 
we take the average over this ensemble.
\begin{figure}
\begin{center}
\includegraphics*[scale=0.35]{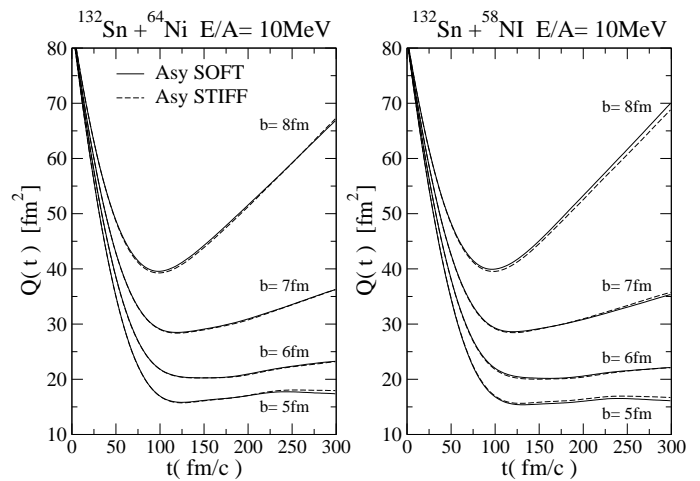}
\end{center}
\caption{Time evolution of the space quadrupole moments for different 
centralities and for the two systems. 
Solid line: Asysoft. 
Dashed line: Asystiff.}
\label{squad}
\end{figure}
\begin{figure}
\begin{center}
\includegraphics*[scale=0.35]{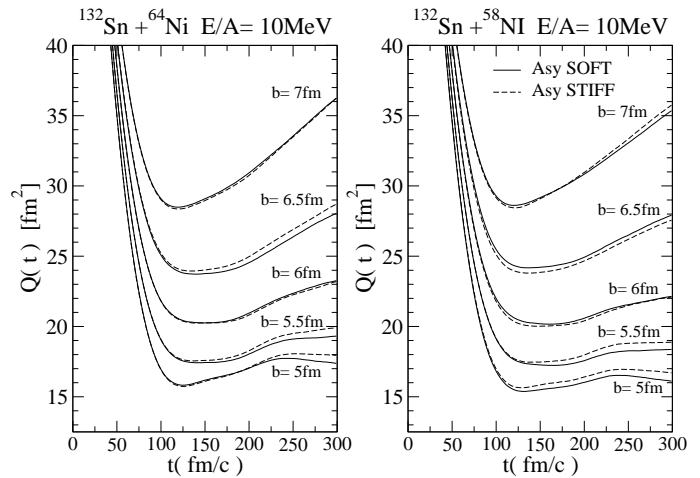}
\end{center}
\caption{Like Fig.\ref{squad} but more detailed in the angular momentum 
transition region, between b=5.0 and 7.0 fm.
Solid line: Asysoft. Dashed line: Asystiff.}
\label{squadlarge}
\end{figure}
In Figs.\ref{squad}, \ref{squadlarge}
we present the time evolution of the  mean space quadrupole moment at various 
centralities for the two reactions and for the two choices of the 
symmetry term. 
We notice the difference in Q(t) between the behavior corresponding to more 
peripheral impact parameters and that obtained for b=5-6 fm, where 
we have still a little oscillation in the time interval between 100 and 300 
fm/c, 
good indication of a fusion contribution. 

We can interpret these observations assuming that starting from about 
b = 5 fm, we 
have a transition from fusion to a break-up mechanism, like deep-inelastic. 
Positive values of the Q(t)-slope should be associated with a 
quadrupole deformation velocity of the dinuclear system that is going 
to a break-up exit channel. We notice a slight systematic difference, 
especially in the 
most neutron-rich system, with a larger deformation velocity in the Asystiff 
case, 
see the more detailed picture of Fig.\ref{squadlarge}.
Hence, 
just from this simple analysis of the average space quadrupole 
``trajectories'' we can 
already appreciate that
the Asysoft choice seems to lead 
to larger fusion cross sections, at least for less peripheral impact 
parameters, between b=5.0 fm and b=6.5 fm.
\begin{figure}
\begin{center}
\includegraphics*[scale=0.35]{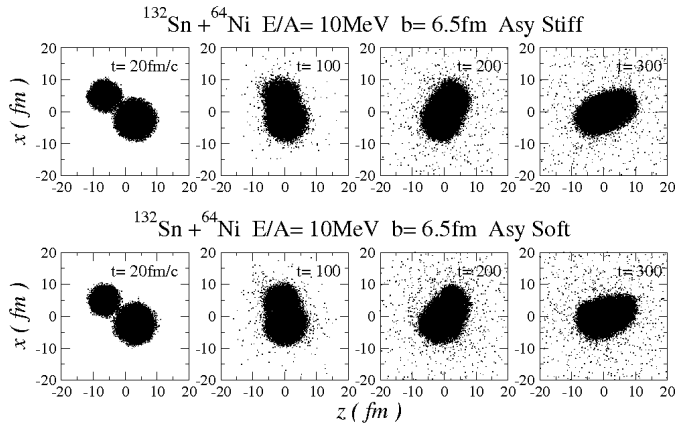}
\end{center}
\caption{Time evolution of the space density distributions for the reaction
$^{132}Sn+^{64}Ni$ (n-rich systems), 10 AMeV beam energy, for semicentral
collisions, b=6.5 fm impact parameter (average over 20 events). 
Upper Panel: Asystiff. Lower Panel: Asysoft. 
}
\label{dens}
\end{figure}
The latter point can also be qualitatively seen from the time evolution
of the space density distributions projected on the reaction plane, as
shown in Fig.\ref{dens}. The formation of a more compact configuration 
in the Asysoft case can be related to a larger fusion probability.

\begin{figure}
\begin{center}
\includegraphics*[scale=0.33]{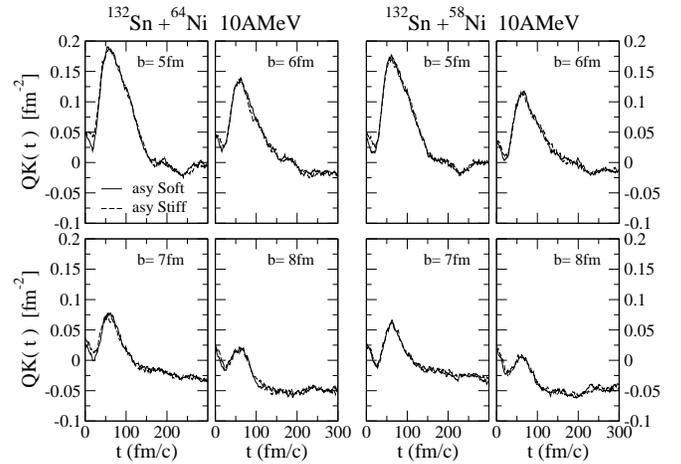}
\end{center}
\caption{Time evolution of the momentum quadrupole moments,
in a sphere of radius 3 fm around the c.o.m., for different 
centralities and for the two systems. 
Solid line: Asysoft. Dashed line: Asystiff.}
\label{pquad}
\end{figure}

It is very instructive to look also at the time
evolution of the quadrupole 
deformations in momentum space. For each event we perform the calculation
in a spherical cell of radius 3 fm around the system center of mass. 
In Fig.\ref{pquad} we 
present the 
time evolution of the average 
p-quadrupole moments at various centralities for 
the two systems and the two choices of the symmetry term.
We  notice a difference between the plots corresponding to  
peripheral or central collisions. 
With increasing impact parameter the quadrupole QK(t) becomes more negative
in the time interval between 100 and 300 fm/c: the components perpendicular 
to the 
symmetry axis, that is rotating in reaction plane, are clearly increasing. 
We can interpret this effect as due to the presence, in the 
considered region, of Coriolis forces 
that are enhanced when the angular momentum is larger. These forces help to 
break 
the deformed dinuclear system. Then the break-up probability will be larger
if the quadrupole moment in p-space is more negative.
From Fig.s 3,5 one can see that there is a region of impact parameter 
(b = 5-6.5 fm)
where the derivative of the quadupole moment in coordinate space, $Q'$, and 
the quadrupole
moment in momentum space, $QK$, are both rather close to zero. This is the 
region where we
expect that fluctuations of these quantities should play an important role 
in determining
the fate of the reaction and event-by-event analysis is essential to 
estimate fusion
vs. break-up probabilities.


\subsection{Analysis of fluctuations and fusion probabilities for 
$^{132}Sn$ induced reactions}


To define a quantitative procedure to fix the event by event fusion
vs break-up probabilities, we undertake an analysis of the
correlation between the two quadrupole moments introduced in the previous 
Section,
in  the  time interval defined before (100-300 fm/c). 
Another important suggestion to look at correlations comes from the very 
weak presence of isospin as well as symmetry energy effects in the separate 
time evolution of the two quadrupole moments, as we can see 
from Figs.\ref{squad},\ref{squadlarge} and Fig.\ref{pquad}.

\begin{figure}
\begin{center}
\includegraphics*[scale=0.33]{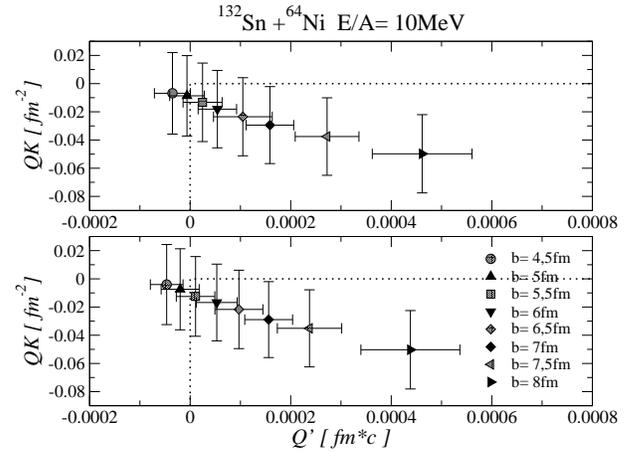}
\end{center}
\caption{$^{132}Sn$ + $^{64}Ni$ system. Mean value and variance of 
QK vs Q', averaged over the 100-300 fm/c 
time interval, at various centralities in the transition region.
The box limited by dotted lines represents the break-up region.
Upper panel: Asystiff. Bottom Panel: Asysoft.}
\label{corr64}
\end{figure}

\begin{figure}
\begin{center}
\includegraphics*[scale=0.33]{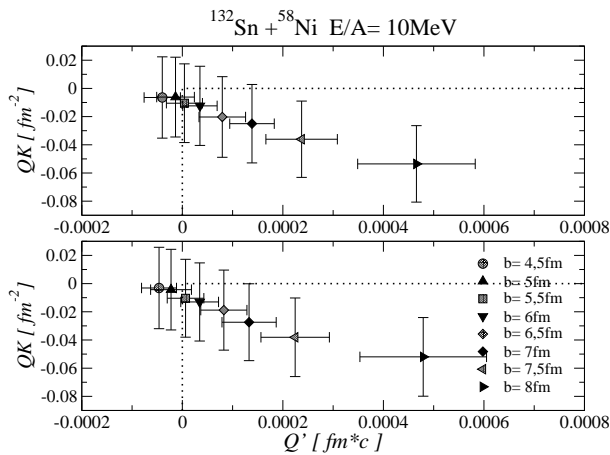}
\end{center}
\caption{Like in Fig.\ref{corr64} but for the $^{132}Sn$ + $^{58}Ni$ 
system.}
\label{corr58}
\end{figure}

Negative $QK$ values denote the presence of velocity components orthogonal
to the symmetry axis, due to angular momentum effects, that help the system 
to separate in two pieces.
At the same time, the observation of a velocity component along the symmetry 
axis indicates that
the Coulomb repulsion is dominating over surface effects (that would  
try to recompact the system), also pushing the system 
in the direction of the break-up.
Hence,
in order to get  the fusion probability from the early evolution of the system 
we assume 
positive Q' and negative $QK$ for break-up events. 
In other words, we suppose that, in the impact parameter range where the 
average value
of the two quantities is close to zero, the system evolution is decided just by
the amplitude of shape fluctuations, taken at the moment when the
formation of a deformed composite system is observed along the SMF 
dynamics (t = 200-300 fm/c, see the contour plots of Fig.4).
Within our prescription, the fusion probability is automatically 
equal to one for central impact parameters, where the system goes
back to the spherical shape and Q' is negative, while it is zero
for peripheral reactions, where Q' is always positive and $QK$ always 
negative.

The correlation plots 
for the two systems studied and the two asy-EOS are represented 
in Figs.\ref{corr64} and \ref{corr58}, respectively.
Through the quantities displayed in the Figures, mean value and variance 
of the two 
extracted properties of the phase space moment evolution, we can evaluate 
the normal curves and the relative areas for each impact parameter 
in order to select the events: break-up events will be located in the regions 
with both positive slope of Q(t) and negative $QK$.
In this way, for each impact parameter we can evaluate the fusion events 
by the 
difference between the total number of events and the number of 
break-up events. Finally the 
fusion cross section is obtained (in absolute value) by
\begin{equation}
\frac{d \sigma}{dl}=\frac{2 \pi}{k^2} l \frac{N_f}{N_{tot}},
\end{equation}
where $l$ is the angular momentum calculated in the semiclassical 
approximation,
 $k$ is the relative momentum of the collision, 
$N_f$ the number of fusion events and $N_{tot}$ the total events in the 
angular 
momentum bin.
\begin{figure}
\begin{center}
\includegraphics*[scale=0.35]{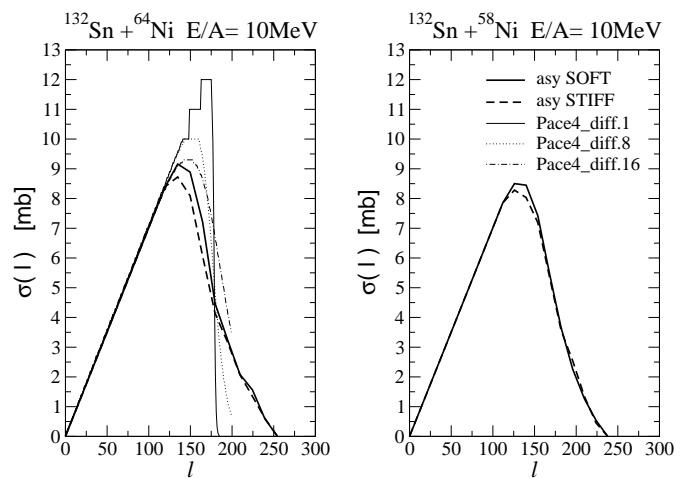}
\end{center}
\caption{Angular momentum distributions of the fusion cross sections (mb) 
for the 
two reactions and the two choices of the symmetry term. For the 
$^{132}Sn$ + $^{64}Ni$ system (left panel),
the results of PACE4 calculations are also reported, for different 
l-diffuseness.}
\label{sigmafus}
\end{figure}
In Fig.\ref{sigmafus} we present the fusion spin distribution plots. 
We note that just 
in the centrality transition region there is a  difference between the 
$\sigma$-fusion corresponding to the two different asy-EOS, with larger values
for Asysoft.

In fact, the total cross sections are very similar: the difference 
in the area is about 4-5 $\%$ in the neutron rich system, $1128~ mb$ (Asysoft)
vs. $1078~ mb$ (Asystiff), and even smaller, $1020~ mb$ vs. $1009~ mb$, 
for the  $^{58}$Ni target. However, through a selection in angular momentum, 
$130 \leq l \leq 180$ ($\hbar$), 
we find that the Asysoft 
curve is significantly above the Asystiff one, and so in this centrality bin 
the fusion cross section difference can reach a 10$\%$ in the case of the 
more neutron-rich system. Then it can be compared to experimental data as 
an evidence of sensitivity to the density dependence of the symmetry energy.

From the comparison of the total areas for the two systems we can also 
estimate  isospin effects on the total fusion cross section, with
a larger value in the more neutron-rich case, as also recently observed
in fusion reactions with Ar + Ni \cite{indrafus} and Ca + Ca isotopes
\cite{Chimerafus}. 
We note that this effect is also, slightly, dependent on the 
symmetry term: The total fusion cross section for the more neutron rich 
system is $10\%$ larger in the Asysoft calculation and about $7\%$
in the Asystiff case.

Finally we like to note that for the neutron-rich case, $^{132}Sn+^{64}Ni$,
our absolute value of the total fusion cross section presents a good agreement
with recent data, at lower energy (around 5 AMeV), taken at the ORNL 
\cite{liang07}.

In Fig.\ref{sigmafus} for the same system (left panel) we show also 
the results obtained with
the macroscopic fusion probability evaluation code $PACE4$,
 \cite{gavron79,tarasov03} obtained with different l-diffuseness parameters,
  fixing, as input parameters, our total fusion cross section and 
maximum angular momentum. 
We see that in order to have a shape more similar to our $\sigma(l)$ 
distribution we have to choose rather large diffuseness values, while
the suggested standard choice for stable systems is around $\Delta l$=4.
This seems to be a nice evidence of the neutron skin effect.

Our main  conclusion is that we can extract significant signals on the 
event by 
event reaction mechanism  by the fluctuations of the quadrupole moments  
in phase space evaluated in a time region well compatible with the interval 
where the transport results are reliable.

\section{Analysis of symmetry energy effects}

The larger fusion probability obtained with the Asysoft choice, 
especially in the more n-rich 
system, seems to indicate that the reaction mechanism is 
regulated by the symmetry term at suprasaturation density, where the 
Asysoft choice is less repulsive for the neutrons \cite{bar05a,col98}.
In order to check this point we have performed a detailed study of the 
density evolution  in the region of overlap of the two nuclei, named 
$neck$ in the following.
\begin{figure}
\begin{center}
\includegraphics*[scale=0.33]{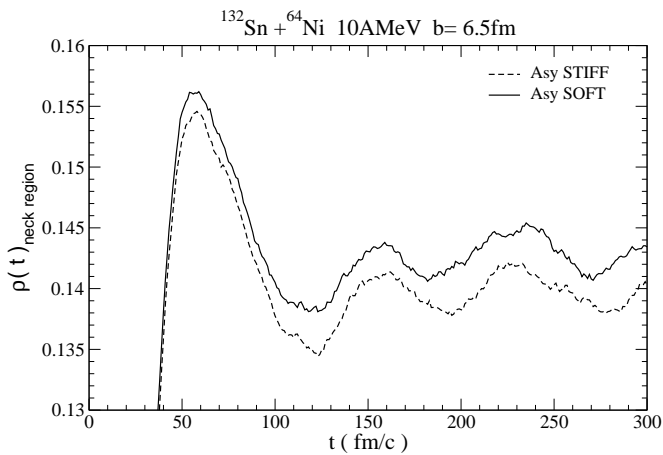}
\end{center}
\caption{Reaction $^{132}Sn+^{64}Ni$ semiperipheral. 
Time evolution of the total density in the ``neck'' region}
\label{rhoneck}
\end{figure}
We present results obtained for
the system  $^{132}Sn+^{64}Ni$
at impact parameter b = 6.5 fm.  
 To account for the system mass asymmetry,
this ``neck'' region is identified 
by a sphere of radius 3 fm 
centered on the symmetry axis, at a distance 
from the projectile center of mass  equal to 
$d(t)*R_1/(R_1+R_2)$, where 
$R_1$ and $R_2$ are the radii of projectile and target, 
and $d(t)$ is the 
distance between the centers of mass of the two colliding nuclei.
In fact, in the time interval of interest for the fusion/break-up dynamics 
it will almost coincide with the system center of mass, see also the 
contour plots of Fig.4. 

The time evolution of the total density in this ``neck'' region is
reported in Fig.\ref{rhoneck} for the two choices of the symmetry energy .
We note that in the time interval of interest
we have densities above or around the normal density and so a less repulsive
symmetry term within the Asysoft choice, corresponding to larger fusion 
probabilities.

This also explains why larger fusion cross sections are seen for the
neutron rich system, mainly in the Asysoft case. 
   In fact, the neutron excess pushes the formed hot compound nucleus 
closer to the
stability valley, especially when the symmetry energy is smaller.

Other nice features are: i) the density values found in the Asysoft 
case are always
above the Asystiff ones, to confirm the expectation of a smaller equilibrium
density for a stiffer symmetry term \cite{bar05a}; ii) collective monopole 
oscillations
are present after 100 fm/c, showing that also at these low energies we can 
have some
compression energy. 

\begin{figure}
\begin{center}
\includegraphics*[scale=0.33]{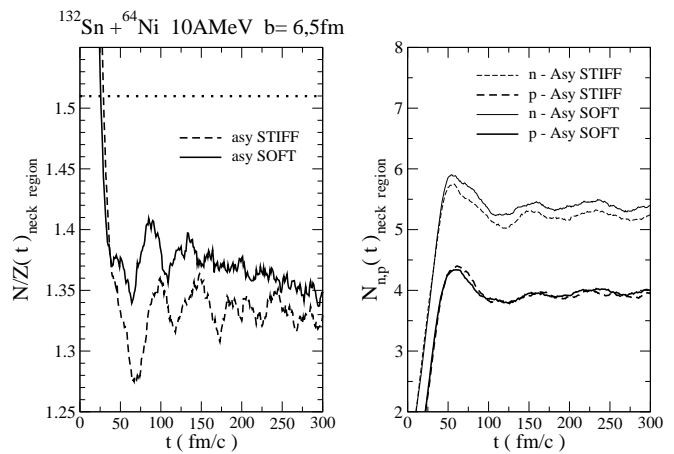}
\end{center}
\caption{Reaction $^{132}Sn+^{64}Ni$ semiperipheral.Left panel: 
time evolution of the neutron/proton ratio in the ``neck'' region.
The dotted line corresponds to the initial isospin asymmetry of the
composite system. Right panel: time evolution of the neutron and proton
densities.}
\label{nzneck}
\end{figure}

It is also instructive to look at the evolution of the isospin content,
the N/Z ratio, in this ``neck'' region, plotted in Fig.\ref{nzneck}.
As reference we show with a dotted line the initial average isospin
asymmetry. 
We see that in the Asysoft choice a systematic larger isospin content
is appearing (Left Panel). This is 
consistent with the presence of a less repulsive neutron potential
at densities just above saturation probed in the first $100fm/c$,
when the fast nucleon emission takes place (Figs.\ref{rhoneck} 
and \ref{nzneck}, Left Panel). 
All that is confirmed by the separate behavior of the
neutron and proton densities shown in the Right Panel of Fig.\ref{nzneck}.

It is finally very interesting the appearance of N/Z oscillations
after 100 fm/c.
This can be related to the excitation of   
isovector density modes in the composite 
system during the path to fusion or break-up. 
Since initially a charge asymmetry is present in the system
(N/Z=1.64 for $^{132}$Sn and 1.28 for $^{64}$Ni)
we expect the presence of collective isovector oscillations during the 
charge equilibration dynamics for $ALL$ dissipative collisions, 
regardless of the final exit channel.
The features of this isovector mode, the Dynamical 
Dipole already observed in
fusion reactions with stable beams \cite{bardip09},
will be further discussed in Section V.

\subsection*{Break-up Events}

Within the same transport approach, a first analysis of symmetry energy
effects on break-up events in semiperipheral collisions of
$^{132}Sn+^{64}Ni$ at $10~AMeV$ has been reported in ref.\cite{NN06}.
Consistently with the more accurate study presented here, smaller
break-up probabilities have been seen in the Asysoft choice.
Moreover the neck dynamics on the way to separation is found also
influenced by the symmetry energy below saturation. This can be
observed in the different deformation pattern of the Projectile-Like 
and Target-Like Fragments (PLF/TLF),
 as shown in Fig.1 of \cite{NN06}. Except for the most peripheral selections,
larger deformations are seen in the Asystiff case, corresponding to a 
smaller symmetry repulsion at the low densities probed in the separation
region. The neutron-rich neck connecting the two partners can then survive 
a longer time producing very deformed primary PLF/TLF. Even small clusters
can be eventually dynamically emitted leading to ternary/quaternary 
fragmentation events \cite{skwira08,wilcz10}.

In conclusion not only the break-up probability but also a detailed study
 of fragment deformations in deep-inelastic (and fast-fission) processes,
as well as of the yield of 3-4 body events, will give independent information
on the symmetry term around saturation.

\section{The Prompt Dipole Mode in Fusion and Break-up Events}

\begin{figure}
\begin{center}
\includegraphics*[scale=0.35]{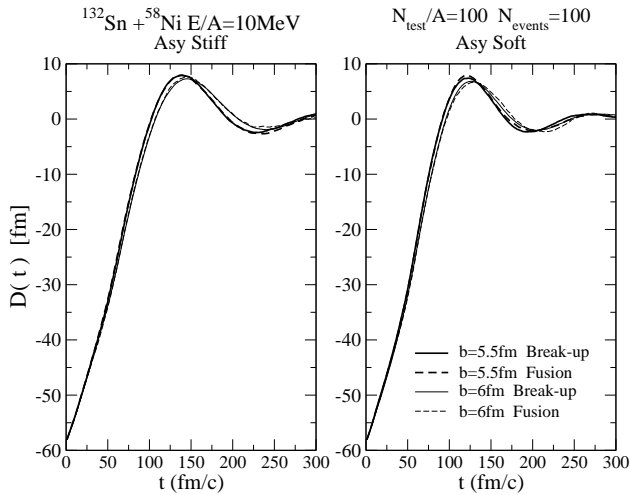}
\end{center}
\caption{Reaction $^{132}Sn+^{58}Ni$ semiperipheral. Prompt Dipole
oscillations in the composite system for break-up (solid lines)
and fusion (dashed lines) events. Left Panel: Asystiff.
Right Panel: Asysoft.}
\label{dipboth}
\end{figure}

\begin{figure}
\begin{center}
\includegraphics*[scale=0.35]{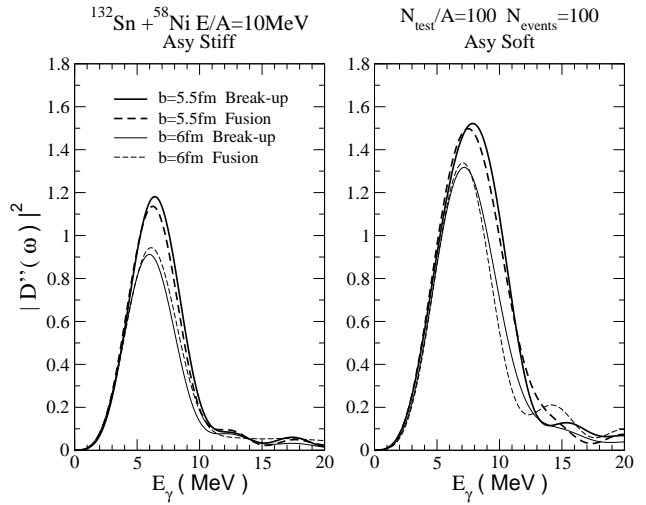}
\end{center}
\caption{Reaction $^{132}Sn+^{58}Ni$ semiperipheral. Prompt Dipole
strengths (in $c^2$ units), see text, for break-up (solid lines)
and fusion (dashed lines) events. Left Panel: Asystiff.
Right Panel: Asysoft.}
\label{strengths}
\end{figure}

\begin{figure}
\begin{center}
\includegraphics*[scale=0.35]{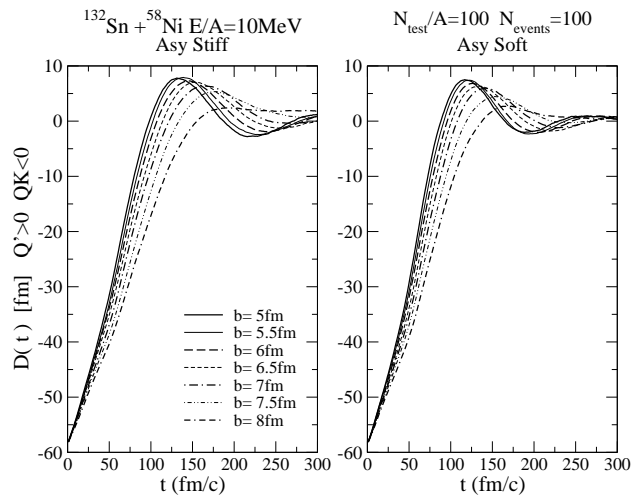}
\end{center}
\caption{Reaction $^{132}Sn+^{58}Ni$ semiperipheral to peripheral.Prompt 
Dipole
oscillations in the composite system for break-up event selections
at each impact parameter. Left Panel: Asystiff.
Right Panel: Asysoft.}
\label{dipbreak}
\end{figure}

From the time evolution
of the nucleon phase space occupation, see Eq.(1), it is possible to extract
at each time step the isovector dipole moment of the
composite system. This is given by
$D(t)=\frac{NZ}{A}X(t)$, where  $A=N+Z$, and $N=N_1+N_2$, $Z=Z_1+Z_2$, are
the total number of participating nucleons, while $X(t)$ is the distance 
between
the centers of mass of protons and neutrons.
It has been clearly shown, in theory as well as in experiments, that
at these beam energies the charge equilibration in fusion reactions 
proceeds through such prompt collective mode. In our study we
have focused the attention on the system with larger initial charge asymmetry,
 the $^{132}$Sn on $^{58}$Ni case,

In Fig.\ref{dipboth} we present the prompt dipole oscillations obtained for
semicentral impact parameters, in the transition zone. We nicely see
that in both classes of events, ending in fusion or deep-inelastic 
channels, the dipole mode is present almost with the same strength.
We note that such fast dipole radiation was actually  observed even in 
the most dissipative deep-inelastic events in stable ion collisions
\cite{PierrouEPJA16,PierrouPRC71,Amorini2004}.

The corresponding emission rates
can be evaluated, 
through a ''bremsstrahlung'' mechanism, in a consistent transport 
approach to the 
rection dynamics, which can account for
the whole contribution along
the dissipative non-equilibrium path, in fusion or deep-inelastic processes
\cite{bar01}. 

In fact from the dipole evolution $D(t)$
we can directly estimate the photon emission probability
($E_{\gamma}= \hbar \omega$):
\begin{equation}
\frac{dP}{dE_{\gamma}}= \frac{2 e^2}{3\pi \hbar c^3 E_{\gamma}}
 |D''(\omega)|^{2}  \label{brems},
\end{equation}
where $D''(\omega)$ is the Fourier transform of the dipole acceleration
$D''(t)$. We remark that in this way it is possible
to evaluate, in {\it absolute} values, the corresponding pre-equilibrium
photon emission yields.

In Fig.\ref{strengths} we report the prompt dipole strengths
$|D''(\omega)|^{2}$ for the same event selections of Fig.\ref{dipboth}.

The dipole strength distributions are very similar in the fusion 
and break-up 
selections in this centrality region where we have a strong competition 
between the two mechanisms. In any case there is a smaller strength 
in the less central collisions 
(b=6.0fm), with a centroid slightly shifted to lower values, 
corresponding to more deformed shapes of the dinuclear composite system.

In the Asysoft choice we have a systematic increase of the yields, 
roughly given by the area of the strength distribution, of about $40\%$
more than in the Asystiff case, for both centralities and selections.
In fact from Eq.(\ref{brems}) we can directly evaluate the total
$\gamma$-multiplicities, integrated over the dynamical dipole region.
For centrality b=5.5fm we get $2.3~10^{-3}$ ($1.6~10^{-3}$) in the
Asysoft (Asystiff) choice, and for b=6.0fm respectively $1.9~10^{-3}$
($1.3~10^{-3}$), with almost no difference betwen fusion and break-up events.

From Fig.\ref{esym} 
we see that Asysoft
corresponds to a larger symmetry energy below saturation. Since the symmetry 
term gives the restoring force of the dipole mode, our result is a good 
indication that the prompt dipole oscillation is taking place in a  
deformed dinuclear composite system, where low density surface contributions
are important, as already observed in ref.\cite{bardip09}.

In the previous Sections we have shown that the Asysoft choice 
leads to a large fusion probability since it gives a smaller repulsion
at the  suprasaturation densities of the first stage of the reaction.
Here we see that for the dipole oscillation it gives a larger restoring force
corresponding to mean densities below saturation. This apparent contradictory
conclusion can be easily understood comparing Figs.\ref{rhoneck} and
\ref{dipboth}. We note that the onset of the collective dipole mode is 
delayed with respect to the first high density stage of the neck region since
the composite system needs some time to develop a collective response of 
the dinuclear mean field.

In this way fusion and dynamical dipole data can be directly used to probe
the isovector part of the in medium effective 
interaction {\it below and above} saturation density.

Another interesting information is derived from Fig.\ref{dipbreak}
where we show the prompt dipole oscillations only for break-up events
at centralities covering the range from semicentral to peripheral.
We nicely see that the collective mode for charge equilibration, due to 
the action of the mean field of the dinuclear system, is disappearing for
the faster, less dissipative break-up collisions.

\subsection*{Anisotropy}

Aside the total gamma spectrum the corresponding
angular distribution can be a sensitive probe to explore the 
properties of preequilibrium dipole mode and the early stages of
fusion dynamics. In fact a clear anisotropy vs. the beam axis
has been recently observed \cite{martin08}.
For a dipole oscillation just along the beam axis we expect an angular 
distribution of the emitted photons like $W(\theta)\sim \sin^2 \theta 
\sim 1+a_2P_2(cos \theta)$ with $a_2=-1$, where $\theta$ is the polar angle
between the photon direction and the beam axis. Such extreme anisotropy
will be never observed since in the collision the prompt dipole axis will 
rotate during the radiative emission. In fact the deviation from the 
$\sin^2 \theta$ behavior will give a measure of the time interval of the fast 
dipole emission. In the case of a large rotation one can even observe 
a minimum at 90 degrees.

\begin{figure}
\begin{center}
\includegraphics*[scale=0.33]{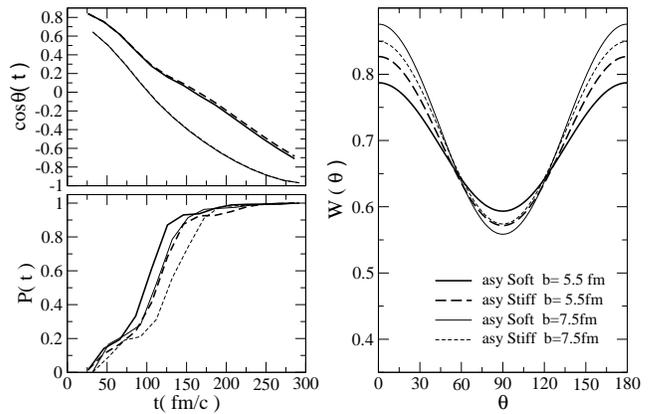}
\end{center}
\caption{reaction $^{132}Sn+^{58}Ni$ semiperipheral.Upper Left panel: 
Rotation angle. Bottom Left Panel: emission probabilities.
Right panel: Weighted angular distribution.
}
\label{anisotropy}
\end{figure}

Let us denote by $\phi_i$ and $\phi_f$ the initial and final angles of the
symmetry axis (which is also oscillation axis) with respect to the beam axis,
associated respectively to excitation and complete damping of the dipole mode.
Then  $\Delta \phi=\phi_f-\phi_i$ is the rotation angle 
during the collective oscillations. We can get the angular distribution in 
this case
 by averaging only over the angle $\Delta \phi$ obtaining

\begin{equation}
W(\theta) \sim 1-(\frac{1}{4}+\frac{3}{4}x)P_2(cos \theta)
\label{angdis}
\end{equation}
where $x=cos (\phi_f$+$\phi_i)\frac{sin (\phi_f-\phi_i)}{\phi_f-\phi_i}$ .    

The point is that meanwhile the emission is damped.  

Within the bremsstrahlung approach we can perform an accurate evaluation 
of the prompt dipole angular distribution using a weighted form where the 
time variation of the radiation emission probability is accounted for
\begin{equation}
W(\theta)=\sum_{i=1}^{t_{max}} \beta_i W(\theta,\Phi_i)
\label{wweighted}
\end{equation}
We divide the dipole emission time in $\Delta t_i$ intervals with the 
corresponding $\Phi_i$ mean rotation angles and the related radiation 
emission probabilities
$\beta_i=P(t_i)-P(t_{i-1})$, where 
$P(t)=\int_{t_0}^{t} \mid D''(t) \mid^2 dt / P_{tot}$
with $P_{tot}$ given by $P(t_{max})$, total emission probability at the
final dynamical dipole damped time.

In Fig.\ref{anisotropy}, upper left panel, we plot the time dependence of 
the rotation angle,
for the $^{132}$Sn + $^{58}$Ni system,  
extracted 
from all the events, fusion and break-up, at two semiperipheral 
impact parameters, for the two symmetry terms. We note that essentially 
the same
curves are obtained with the two Iso-EoS choices: the overall rotation 
is mostly ruled by
the dominant isoscalar interaction.  

Symmetry energy effects will be induced by the different time evolution 
of the emission probabilities, as shown in the bottom left panel.


We clearly see that the dominant 
emission region is the initial one, just after the onset of the collective 
mode between 80 and 150 fm/c, while the 
emitting dinuclear system has a large rotation. 
Another interesting point is the dependence on the symmetry energy. 
With a weaker symmetry term at low densities (Asystiff case), 
 the $P(t)$ is
a little delayed and presents a smoother behavior. As a consequence,
according to Eq.(7), we can 
expect possible symmetry energy effects even on the angular distributions.

This is shown in the right panel of Fig.\ref{anisotropy}, where we have the 
weighted distributions
 (Eq.(\ref{wweighted})), for the two impact parameters and
 the two choices of the symmetry 
energies. We see some sensitivity to the stiffness of the symmetry term.
Hence, from accurate measurements of the angular distribution of
the emitted $\gamma$'s, in the range of impact parameters 
where the system rotation is significant, one can extract  
independent information  on the density behavior of the symmetry energy.

\section{Conclusions and perspectives}
We have undertaken an analysis of the reaction path followed in collisions 
involving exotic systems at beam energies around 10 AMeV. 
In this energy regime, the main reaction mechanisms range from fusion to
dissipative binary processes, together with the excitation of collective
modes of the nuclear shape. In reactions with exotic systems, these
mechanisms are expected to be sensitive to the isovector part of the 
nuclear interaction, yielding information on the density dependence of 
the symmetry energy. Moreover, in charge asymmetric systems, isovector dipole
oscillations can be excited at the early dynamical stage, also sensitive
to the behavior of the symmetry energy. 
We have shown that, in neutron-rich systems, fusion vs. break-up 
probabilities are influenced by the neutron repulsion during the approaching
phase, where densities just above the normal value are observed. 
Hence larger fusion cross sections are obtained in the Asysoft case, associated
with a smaller value of the symmetry energy at supra-saturation densities. 
On the other hand, the isovector collective response, that takes place in the 
deformed dinuclear configuration with large surface contributions, is sensitive
to the symmetry energy below saturation.
 
The relevant point of our analysis is that it is based just on the study of
the fluctuations that develop during the early dynamics, when the transport
calculations are reliable. Fluctuations of the quadrupole moments, in phase 
space, essentially determine the final reaction path. 
It should be noticed that the fluctuations discussed here are 
essentially of thermal nature.
It would be interesting to include also the contribution
of quantal (zero-point) fluctuations of surface modes and angular momentum.
Indeed the frequencies of the associated collective motions are 
comparable to 
the temperature ($T\approx 4~ MeV$) reached in our reactions \cite{Lan_book}. 
This would increase the
overall amplitude of surface oscillations, inducing larger fluctuations 
in the system
configuration and a larger break-up probability.
Such quantum effect has been recently shown to be rather important for fusion
probabilities at near and sub-barrier energies \cite{ayik10}.
The agreement of our semiclassical procedure with present data above the 
barrier could be an indication of a dominance of thermal fluctuations 
at higher excitation energy. In any case this point should be more carefully
studied.
 
Finally, we would like to stress that, according to our analysis,
considerable isospin effects are revealed just selecting the impact
parameter window corresponding to semi-peripheral reactions. 
Interesting perspectives are opening for new experiments on low energy 
collisions with exotic beams focused to the study of the symmetry term 
below and above
saturation density. We suggest some sensitive observables:

i) Fusion vs. Break-up probabilities in the centrality transition region;

ii) Fragment deformations in break-up processes and probability of
ternary/quaternary events.

iii) $\gamma$-multiplicity and anisotropy of the Prompt Dipole Radiation,
 for dissipative collisions in charge asymmetric entrance channels.

\vskip 0.3cm
\noindent

{\bf Aknowledgements}

We warmly thank Alessia Di Pietro for the discussions about the
the use of Pace4 fusion simulations. 
One of authors, V. B. thanks for hospitality at the Laboratori
Nazionali del Sud, INFN. This work was supported in part by the Romanian
Ministery for Education and Research under the contracts PNII, No.
ID-946/2007 and ID-1038/2008.

\end{document}